%% file: sigmatolamee_muniscript.tex
\documentclass[prd,aps,reprint,noeprint,superscriptaddress,amsmath,amssymb]{revtex4-1}

\usepackage[compat=1.1.0]{tikz-feynman}
\usepackage{graphicx}
\usepackage{subfigure}
\usepackage[english]{babel}
\usepackage{epsfig}
\usepackage{multirow}
\usepackage{overpic}
\usepackage{color}
\usepackage{bm}
\usepackage{xspace}
\usepackage{booktabs,tabularx}
\usepackage{array}
\usepackage{textcomp}
\usepackage[colorlinks,linkcolor=blue,urlcolor=blue,citecolor=blue]{hyperref}
\usepackage{adjustbox}
\usepackage{array}
\usepackage{ragged2e}
\usepackage[utf8]{inputenc}

\usepackage{cleveref}

\newcommand{\Jpsisigsigbar}  {J/\psi\to\Sigma^{0}\bar{\Sigma}^{0}}
\newcommand{\Siglamee}  {\Sigma^{0}\to\Lambda e^{+}e^{-} }

\newcommand{\Sigbarlambaree}  {\bar{\Sigma}^0 \to\bar{\Lambda} e^{+}e^{-} }

\newcommand{\Sigbarlambargam} {\bar{\Sigma}^{0}\to\bar{\Lambda}\gamma}
\newcommand{\Lamppim} {\Lambda\to p\pi^{-}}

\newcommand{\Sigmab}{\bar{\Sigma}^0}
\newcommand{\Sigmaa}{\Sigma^0}
\newcommand{\Lambdab}{\bar{\Lambda}}
\newcommand{\ee}{e^{+}e^{-}}

\newcommand{\Jpsirhopi}  {J/\psi\to\rho^{0}\pi^{0}}

\newcommand{\jpsi}{J/\psi}

\newcommand{\pip}{\pi^+}
\newcommand{\pim}{\pi^-}

\newcommand{\bfg}{\begin{figure}}
\newcommand{\efg}{\end{figure}}
\newcommand{\bitm}{\begin{itemize}}
\newcommand{\eitm}{\end{itemize}}
\newcommand{\bnum}{\begin{enumerate}}
\newcommand{\enum}{\end{enumerate}}
\newcommand{\btbl}{\begin{table}}
\newcommand{\etbl}{\end{table}}
\newcommand{\btbu}{\begin{tabular}}
\newcommand{\etbu}{\end{tabular}}

\newcommand{\beq}{\begin{equation}}
\newcommand{\edq}{\end{equation}}

\newcommand{\BESIIIorcid}[1]{\href{https://orcid.org/#1}{\hspace*{0.1em}\raisebox{-0.45ex}{\includegraphics[width=1em]{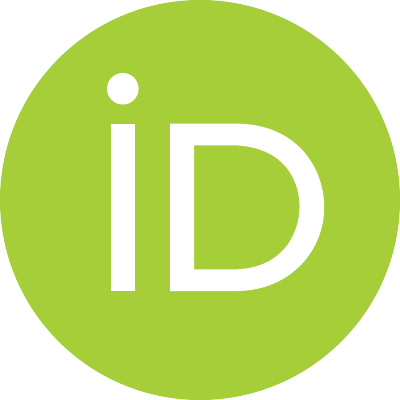}}}}

\begin{document}
\normalsize
\parskip=5pt plus 1pt minus 1pt

\title{
\boldmath  Measurement of Branching Fraction and Transition Magnetic Moment of the Hyperon Dalitz Decay $\Sigma^0 \rightarrow \Lambda e^+e^-$
}

\author{\input{./authorlist_2025-11-14.tex}\mbox{}}

\begin{abstract}
Based on a data sample of 10 billion  $J/\psi$ events collected with the BESIII detector operating at the BEPCII collider, the Dalitz decay $\Siglamee$ is studied experimentally for the first time.   The $\Sigma^0$ hyperons are produced through the process $\Jpsisigsigbar$ and analyzed using a double tag method. The absolute branching fraction is measured to be $ \mathcal{B}(\Siglamee) = (6.34 \pm 0.25_{\textrm{stat.}} \pm 0.23_{\textrm{syst.}}) \times 10^{-3} $. This result shows a $2\sigma$ discrepancy from the theoretical calculation quoted in the PDG, where the uncertainties are statistical and systematic, respectively. In addition to the branching fraction, the transition magnetic moment $\mu$ is determined to be $(1.74 \pm 0.03_{\textrm{stat.}} \pm 0.09_{\textrm{syst.}}) \, \mu_{N}$, where $\mu_N = \frac{e}{2m_p}$ represents the nucleon magnetic moment, providing valuable insight into the intrinsic structure of the $\Sigmaa$ hyperon.
\end{abstract}

\maketitle

The neutral member of the $\Sigma$ hyperon isotriplet, the $\Sigmaa$, was first observed using a propane bubble chamber in 1956~\cite{Plano1957}.
Among the SU(3) hyperon ground states, the $\Sigmaa$ is the only one that decays dominantly through an electromagnetic (EM) interaction~\cite{Particledatagroup}.  It therefore plays a unique role in revealing the intrinsic structure of hyperons and the underlying interactions between photons and hyperons.
Notably, the EM Dalitz decay $\Sigmaa\to\Lambda \ee$, predicted in 1958~\cite{Feinberg1958}, offers a powerful probe into the EM structure of the $\Sigmaa-\Lambda$ transition.
In this decay, the $\ee$ pair is produced from the internal conversion of an intermediate virtual photon with an invariant mass $M_{\ee}$.
The corresponding decay rate as a function of $M_{\ee}$ can be described precisely by quantum electrodynamics (QED), and the branching fraction (BF) is predicted to be $5.5\times10^{-3}$ through QED calculations~\cite{Feinberg1958, Abers1977, Granados2017, Husek2020}.

The uniqueness of the $\Sigmaa-\Lambda$ transition lies in its ability to provide access to the form factor across a broad energy range, which is unattainable with other ground-state hyperons. Theoretically,  the transition vertex between the $\Sigmaa$ and $\Lambda$ can be described with a transition form factor (FF), which provides insight into the quark-gluon structure of hyperons. Exploring the FF of hyperons is critical to understand their intrinsic structure.

\begin{figure}[t]
	\centering
	{
		\begin{minipage}{1.0\linewidth}
			\includegraphics[scale=0.5]{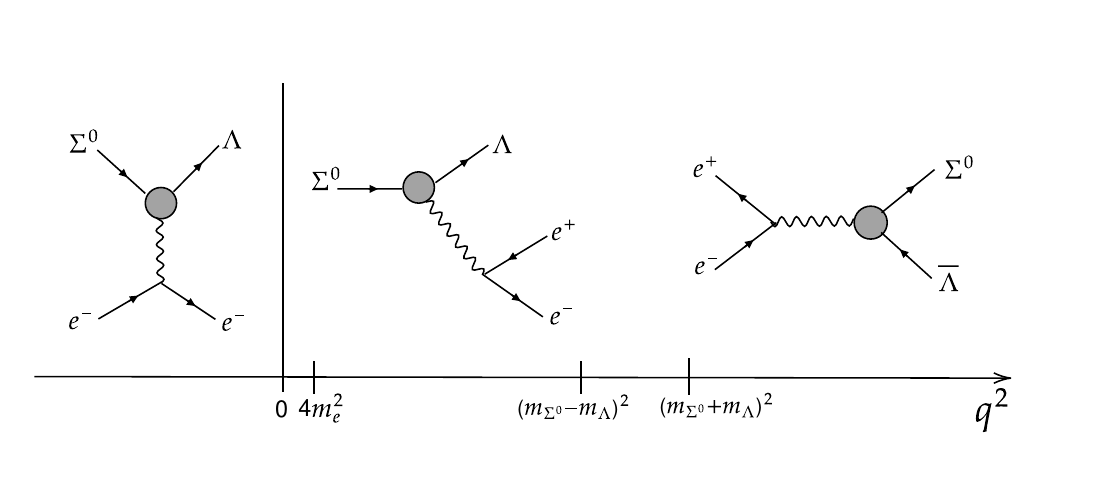}
		\end{minipage}
	}
	\caption{Feynman diagrams showing the $\Sigmaa\Lambda$ interaction vertex in different $q^2$ regions:
		$q^2\textless0$, $4m_{e}^2\textless q^2\textless (M_{\Sigmaa}-M_{\Lambda})^2$ and
		$q^2 \textgreater(M_{\Sigmaa}+M_{\Lambda})^2$ from left to right.}
	\label{fig::feynman}
\end{figure}

Experimentally, the FF can be measured with various processes in their allowed kinematical regions.
As shown in Fig.~\ref{fig::feynman}, the scattering process $e^{-}\Sigmaa\leftrightarrow e^{-}\Lambda$ probes the space-like region ($q^2<0$) while the electron-positron annihilation process $e^{+}e^{-}\rightarrow\Lambda\Sigmab$ covers the time-like region at high $q^2$ ($q^2 \ge (M_{\Sigmaa}+M_{\Lambda})^2$), where $q^2$ is the four-momentum transfer. The EM Dalitz
decay $\Sigma^0 \rightarrow \Lambda e^+e^-$ probes the time-like region at
low $q^2$, namely $4m_e^2 \textless q^2 \textless (M_{\Sigma^0} - M_{\Lambda})^2$. The EM Dalitz decay thereby
fills the gap between the regions accessed by $e^-$ scattering and $e^+e^-$ annihilation.

The transition magnetic moment is a crucial physical quantity for the precise description of transition form factors.  The magnetic moment of the $\Sigmaa$--$\Lambda$ transition was measured in previous experiments in the 1980s~\cite{TFF1,TFF2}, primarily through the scattering of a high-energy $\Lambda$ beam on a fixed target. Hyperon--electron scattering, the traditional method of directly measuring transition form factors in the space-like region, is not possible at an $e^+e^-$ facility such as BESIII. In the time-like region, BESIII has already conducted measurements of this transition at high $q^2$~\cite{BESTFF1,BESTFF2}. However, to this day, experimental results in the low $q^2$ region remain absent, apart from the observation of the $\Siglamee$ Dalitz decay used in the determination of the relative parity of the $\Sigmaa$ and $\Lambda$~\cite{RP1,RP2}. In this Letter, we present a study of the Dalitz decay $\Siglamee$ based on $(10 087 \pm 44) \times 10^{6}$ $J/\psi$ events~\cite{JpsiNumber} collected by the BESIII experiment. The absolute BF of $\Siglamee$ and the transition magnetic moment in the time-like region at low $q^2$ region are determined. Throughout this Letter, the charge conjugate mode $\Sigbarlambaree$ is always included unless otherwise stated.

The BESIII detector~\cite{BESIII} records symmetric $e^+e^-$ collisions provided by the BEPCII storage ring~\cite{BEPCII}, which operates over a center-of-mass energy ($\sqrt{s}$) range from 1.84 to 4.95~GeV and has achieved a peak luminosity of $1.1\times10^{33}$~cm$^{-2}$s$^{-1}$ at $\sqrt{s}=3.770$~GeV.
The cylindrical core of the BESIII detector covers 93\% of the full solid angle and consists of a helium-based multilayer drift chamber~(MDC), a plastic scintillator time-of-flight system~(TOF), and a CsI(Tl) electromagnetic calorimeter~(EMC). All the sub-detectors are enclosed in a superconducting solenoidal magnet providing a 1.0~T magnetic field.  The magnetic field was 0.9~T in 2012, which affects 10.8\% of the total $J/\psi$ data. The solenoid is supported by an octagonal flux-return yoke with resistive plate counter muon identification modules interleaved with steel. The charged-particle momentum resolution at $1~{\rm GeV}/c$ is $0.5\%$, and the d$E$/d$x$ resolution is $6\%$ for electrons from Bhabha scattering. The EMC measures photon energies with a resolution of $2.5\%$ ($5\%$) at $1$~GeV in the barrel (end cap) region. The time resolution in the plastic scintillator TOF barrel region is 68~ps, while that in the end cap region was 110~ps. The end cap TOF system was upgraded in 2015 using multi-gap resistive plate chamber technology, providing a time resolution of 60~ps~\cite{etof}, which benefits 87.0\% of the data used in this analysis ~\cite{Li2017, etof, CAO2020163053}.

To perform an absolute BF measurement of $\Siglamee$ and to improve the signal-to-background ratio, a double tag (DT) approach is utilized in this analysis.
Candidates of $\Jpsisigsigbar$ are first selected by reconstructing a $\Sigmab$ in its dominant decay mode $\Sigbarlambargam$ with a BF $\mathcal{B}(\Sigbarlambargam) \sim 100\%$~\cite{Particledatagroup}.  These events are denoted as single tag (ST) signal events.
Then the signal decay of $\Siglamee$ is reconstructed from the tracks recoiling against the ST $\Sigmab$.  These events are referred to as DT signal events. In this analysis, the $\Lambda$ is always reconstructed via its decay channel $\Lambda \rightarrow p\pi^-$.
The absolute BF of $\Siglamee$ is then calculated by
\begin{equation}
	\mathcal{B}(\Siglamee)= \frac{N_{\textrm{DT}}} {N_{\textrm{ST}}} \times \frac{\varepsilon_{\textrm{ST}}} {\varepsilon_{\textrm{DT}}} \times \frac{1}{\mathcal{B}(\Lamppim)},
	\label{BF-Siglamee}
\end{equation}
where $N_{\textrm{ST}(\textrm{DT})}$ and $\varepsilon_{\textrm{ST}(\textrm{DT})}$ are the ST (DT) yields and the corresponding detection efficiencies, respectively. $\mathcal{B}(\Lamppim)$ is the decay BF of $\Lamppim$.

Monte Carlo (MC) simulated events are used to determine the detection efficiency, optimize event selection criteria, and study possible backgrounds. GEANT4-based~\cite{Agostinelli2003,Allison2006} MC simulation software, which includes the geometric and material descriptions of the BESIII detector, the time-dependent detector response and beam-related backgrounds, as well as the detector running conditions and performance, is used to generate MC samples. The effects of beam energy spread and initial state radiation in the $e^+e^-$ annihilation are incorporated  with the generator {\sc kkmc}~\cite{kkmc}. The inclusive MC sample includes the production of the $J/\psi$ resonance and the continuum processes incorporated in {\sc kkmc}. The known decay modes are modeled with {\sc evtgen}~\cite{LANGE2001152,PingRongGang2008} with the BFs taken from the PDG~\cite{Particledatagroup}, while the remaining unknown decays are modelled with {\sc lundcharm}~\cite{Yang_2014}. Final-state radiation from charged particles is incorporated using the {\sc photos} package~\cite{PHOTOS}.

Two signal MC samples with different decay topologies, $J/\psi \rightarrow \Sigmaa(\rightarrow$ anything$)\Sigmab(\rightarrow \Lambdab\gamma)$ and $J/\psi \rightarrow \Sigmaa(\rightarrow \Lambda \ee)\Sigmab(\rightarrow \Lambdab\gamma)$, as well as the MC sample for the dominant background process $J/\psi \rightarrow \Sigmaa(\rightarrow \Lambda\gamma)\Sigmab(\rightarrow \Lambdab\gamma)$, are generated according to their helicity amplitudes, where $\Lambda$ always decays into $p\pi^-$.
In the simulation of the signal processes, the helicity amplitudes are described by the eleven observed angles $\xi=({\theta_{\Sigma}, \theta_{\Lambda}, \phi_{\Lambda}, \theta_{\Lambdab}, \phi_{\Lambdab}, \theta_{p}, \phi_{p}, \theta_{\bar{p}}, \phi_{\bar{p}},\theta_{l}, \phi_{l}})$, where $\theta_{\Sigma}$ is the polar angle of the $\Sigmaa$ in the rest frame of the $J/\psi$, and ${\theta_{\Lambda}, \theta_{\Lambdab}, \theta_{p}, \theta_{\bar{p}}, \theta_{l}}$ and ${\phi_{\Lambda}, \phi_{\Lambdab}, \phi_{p}, \phi_{\bar{p}}, \phi_{l}}$ are the polar and azimuthal angles of the daughter particles in the rest frame of their mother particles, respectively. The corresponding differential amplitude is given as $d\sigma \propto \mathcal{W}(\xi)d\xi$.  Here, $\mathcal{W}(\xi)$ is the amplitude density function for the whole decay chain, including the Dalitz decay $\Siglamee$, and can be described by:

\begin{equation}
	\mathcal{W}=\frac{\alpha^2_{em}}{q^4}(q^2-4m_e^2)\sum_{\mu, \bar{v}=0}^3 \sum_{\mu^\prime=0}^3 \sum_{\bar{v}^\prime=0}^3 C_{\mu \bar{\nu}} a_{\mu \mu^\prime}^{\Sigmaa} a_{\mu^\prime 0}^{\Lambda} a_{\bar{\nu}\bar{\nu}^{\prime}}^{\Sigmab} a_{\bar{\nu}^{\prime}0}^{\bar{\Lambda}},
	\label{eq::amplitude}
\end{equation}
where $\alpha_{em}=4\pi/137$ and $m_e$ is the mass of the electron,  $C_{\mu \bar{\nu}}$ is a matrix for the polarization and spin correlation of $\Sigmaa$ and $\Sigmab$, and other matrices like $a_{\mu \mu \prime}^{\Sigmaa}$ are the hyperon decay matrices depending on their production. All these matrices can be found in Ref.~\cite{Batozskaya2023}. The parameters of the helicity amplitude models in the MC simulation are taken from Refs.~\cite{alphajpsi,Deltajpsi}.

The following selection criteria are applied to select signal events: charged tracks must fulfil $|\cos \theta| < 0.93$ and $|V_{z}| < 30$~cm, where $\theta$ is defined with respect to the $z$ axis, which is the symmetry axis of the MDC, and $|V_{z}|$ is the distance from the closest point of the charged track to the interaction point (IP) along the beam direction. The distance from the closest point of the charged track to the IP but perpendicular to the beam direction, $|V_{r}|$, is not restricted due to the long decay length of the $\Lambda$. The total number of good charged tracks in the ST analysis must be at least 2.
Since MC simulation indicates that the $\bar{p}$ and $\pi^+$ from the $\Lambdab$ decay are well separated kinematically, a negatively charged track with momentum larger than 0.55~GeV$/c$ is assigned to be a $\bar{p}$, and that with positive charge and momentum less than 0.35~GeV$/c$ is assigned to be a $\pi^+$.
To further improve the purity of the sample, particle identification (PID) combining the specific ionization energy loss $dE/dx$ from the MDC and the flight time information from the TOF is carried out to calculate the probabilities Prob($h$) ($h=p,K,\pi$) with different particle type assumptions, and the proton candidate is required to satisfy $\text{Prob}(p) \textgreater \text{Prob}(K,\pi)$. No additional PID requirement is applied to the pion candidate.
Each candidate event is required to have at least one $\bar{p}$ and $\pi^+$.

Photon candidates are identified using isolated showers in the EMC. The deposited energy of each shower must be more than 25~MeV in the barrel region ($|\cos\theta| \textless 0.80$) and more than 50~MeV in the end-cap region ($0.86 \textless |\cos\theta| \textless 0.92$). To exclude showers originating from charged tracks, the angle subtended by the EMC shower and the position of the closest charged track at the EMC must be greater than 10$^\circ$ as measured from the IP. To suppress electronic noise and showers unrelated to the collision event, the difference between the EMC time and the event start time is required to be within [0, 700]\,ns.

Based on the selected $\bar{p}$ and $\pi^+$, a $\Lambdab$ candidate is reconstructed by performing vertex and secondary vertex fits to all possible $\bar{p}$ and $\pi^+$ combinations, and the one with the minimum $\chi^2$ of the secondary vertex fit ($\chi^2_{\textrm{sec}}$) and satisfying $\chi^2_{\textrm{sec}} \textless 200$  is kept for further analysis.
Further requirements of $L_{\textrm{decay}} > 0$ and $|M_{\bar{p}\pi^+}-M_{\Lambda}| < 8$~MeV$/c^2$ are used to eliminate the combinatorial background and to improve the purity, where $L_{\textrm{decay}}$ is the $\Lambdab$ decay length obtained from the vertex and secondary vertex fits, $M_{\bar{p}\pi^+}$ is the ${\bar{p}\pi^+}$ invariant mass and $M_{\Lambda}$ is the nominal $\Lambda$ mass taken from the PDG~\cite{Particledatagroup}.

The ST $\Sigmab$ signal is then reconstructed by combining the $\Lambdab$ and a selected photon candidate, where all photon candidates are tried, and the one resulting in $M_{\Lambdab\gamma}^{\textrm{corr}}$  closest to the nominal $\Sigmab$ mass  $M_{\Sigmaa}$ taken from the PDG and satisfying $|M_{\Lambdab\gamma}^{\textrm{corr}}-M_{\Sigmaa}|\textless 15$~MeV$/c^2$  is kept for further analysis, where $M_{\Lambdab\gamma}^{\textrm{corr}}=M_{\Lambdab\gamma}-M_{\bar{p}\pi^+}+M_{\Lambda}$ and $M_{\Lambdab\gamma}$ is the invariant mass of the reconstructed $\Lambdab$ and the selected photon.
To eliminate backgrounds with a $\Sigmab$ in the final state, such as $\jpsi\to\Lambda\Sigmab$ and $\jpsi\to\Sigma^*\Sigmab$,
the mass recoiling against the reconstructed $\Sigmab$ candidate,
\begin{equation}
	M_{\textrm{rec}} = \sqrt{(E_{\textrm{cms}}-E_{\Lambdab}-E_{\gamma})^2-(\textbf{P}_{\Lambdab}+\textbf{P}_{\gamma})^2},
	\label{Mrec_sigma0bar}
\end{equation}
is used to identify the $\Jpsisigsigbar$ signal, where $E_{\textrm{cms}}$ is the center-of-mass energy, $E_{\Lambdab}$ ($\textbf{P}_{\Lambdab}$) and $E_{\gamma}$ ($\textbf{P}_{\gamma}$) are the energy and momentum vector of the $\Lambdab$ and the selected $\gamma$ in the rest frame of the $J/\psi$, respectively.
A detailed study based on the signal MC simulation indicates there are correlations between $M_{\textrm{rec}}$ and $M_{\Lambdab\gamma}$,
therefore the corrected recoiling mass $M_{\textrm{rec}}^{\textrm{corr}}=M_{\textrm{rec}}+M_{\Lambdab\gamma}-M_{\Sigma^0}$ is used in the following analysis.

The resulting distributions of $M_{\Lambdab\gamma}^{\textrm{corr}}$ and  $M_{\textrm{rec}}^{\textrm{corr}}$ for data and the inclusive MC sample with the above selection criteria applied are shown in the Supplemental Material~\cite{supplement}. Good agreement between data and the inclusive MC sample in the signal region of $\Sigmaa\Sigmab$ is observed.
Detailed studies based on the signal MC sample indicate that a small fraction of surviving signal events contains fake photons, defined as those for which the angle between the true photon and the selected one satisfies $\theta_{\textrm{match}} > 5^\circ$.
The background events, including the fake photon, do not peak in the $M_{\textrm{rec}}^{\textrm{corr}}$ distribution.

To extract the ST yields, maximum likelihood (ML) fits are performed to the $M_{\textrm{rec}}^{\textrm{corr}}$ distribution within the range between 1.13 and 1.26 GeV$/c^2$ for the two charge conjugate modes individually.
In the fit, the signal is described with an MC simulated shape convolved with a Gaussian function, which accounts for the resolution difference between data and MC simulation. The background is described with a second-order Chebyshev polynomial function.
The resulting fit curves are shown in the Supplemental Material~\cite{supplement}. The extracted ST signal yields and the detection efficiencies evaluated with the signal MC samples are summarized in Table~\ref{table::Measurement}, where the detection efficiencies are the number of events after applying all selection criteria (including the requirement of $\theta_{\textrm{match}} \textless 5^\circ$) divided by the number of generated events.

\begin{table*}[!htpb]
	\renewcommand\arraystretch{1.2}
    \centering
    \caption{
    Summary of $N_{\textrm{ST}(\textrm{DT})}$ and $\varepsilon_{\textrm{ST}(\textrm{DT})}$, BFs and transition magnetic moments ($\mu$) for the two charge conjugated modes after applying correction factors~\cite{supplement}. The first uncertainties are statistical and the second systematic.
    }
	\label{table::Measurement}
    \scalebox{1.0}
    {
	\begin{tabular}{l c c}
		\hline	\hline
        \noalign{\vskip 3pt}
		Mode                  &  $\Siglamee$ & $\Sigbarlambaree$  \\
		\hline
		ST yield~($\times 10^3$)                  &   $2916.9 \pm 2.7$      &   $3225.2 \pm 2.6$                   \\
		ST efficiency~(\%)          &   36.46                  &   38.13                                \\
		DT yield                  &   348 $\pm$ 23    &   405 $\pm$ 20                   \\
		DT efficiency~(\%)          &   1.04                   &   1.12                                 \\
		Correction factor           &   0.97                   &   0.94                     \\
		\hline
		Individual $\mathcal{B}_{\Siglamee}$($\times 10^{-3}$)   &   $6.38 \pm 0.44 \pm 0.25$   &   $6.31 \pm 0.33 \pm 0.22$          \\
		Simultaneous $\mathcal{B}_{\Siglamee}$($\times 10^{-3}$)    &    \multicolumn{2}{c}{$6.34 \pm 0.25 \pm 0.23$}        \\
		Theoretical Prediction~($\times 10^{-3}$) & \multicolumn{2}{c}{5.5}   \\
		\hline
		Individual $\mu/\mu_{N}$    &   $1.75 \pm 0.07 \pm 0.09$   &   $1.73 \pm 0.05 \pm 0.09$          \\
		Simultaneous $\mu/\mu_N$     &   \multicolumn{2}{c}{$1.74 \pm 0.03 \pm 0.09$}        \\
		\hline \hline
	\end{tabular}
    }
\end{table*}

The signal events $\Siglamee$ are selected in the remaining charged tracks recoiling against the reconstructed ST $\Sigmab$ candidates.
The signal $\Lambda$ is reconstructed using the $p\pi^-$ combination with the same charged track selection criteria and $\Lambda$ reconstruction approach as those in the ST analysis. The difference is that in the DT analysis at least 6 good charged tracks are required.
The $\ee$ pair is selected from the remaining charged tracks apart from $p\bar{p}\pi^+\pi^-$ for the $\Lambda\Lambdab$ candidate, where $e^\pm$ is required to have a vertex satisfying  $|V_{z}| < 10$~cm and $|V_{r}| < 1$~cm, and with the PID requirement $\frac{Prob(e)}{Prob(e)+Prob(\pi)+Prob(K)} > 0.8$.

To improve the momentum and mass resolution of the $e^+e^-$ pair, a vertex fit is performed to the $\ee$ candidate, and the resulting kinematics and vertex position will be used in the following analysis.
A kinematic fit imposing energy and momentum conservation, hereafter denoted as the 4C kinematic fit, is performed under the hypothesis $J/\psi \rightarrow p\bar{p}\pi^+\pi^-\gamma\ee$, where the input kinematics of the $\ee$, $p\pim$ and $\bar{p}\pip$ are from the above $\ee$ pair, $\Lambda$ and $\Lambdab$ vertex fits, respectively.
Each event may have multiple $\ee$ pair candidates. The 4C kinematic fit is performed for all candidates and the one with the minimum $\chi^2_{\textrm{4C}}$ and satisfying $\chi^2_{\textrm{4C}}<200$ is kept for further analysis.

\begin{figure}[htpb]
	\begin{center}
	{
		\begin{minipage}{1.0\linewidth}
			\includegraphics[scale=0.3]{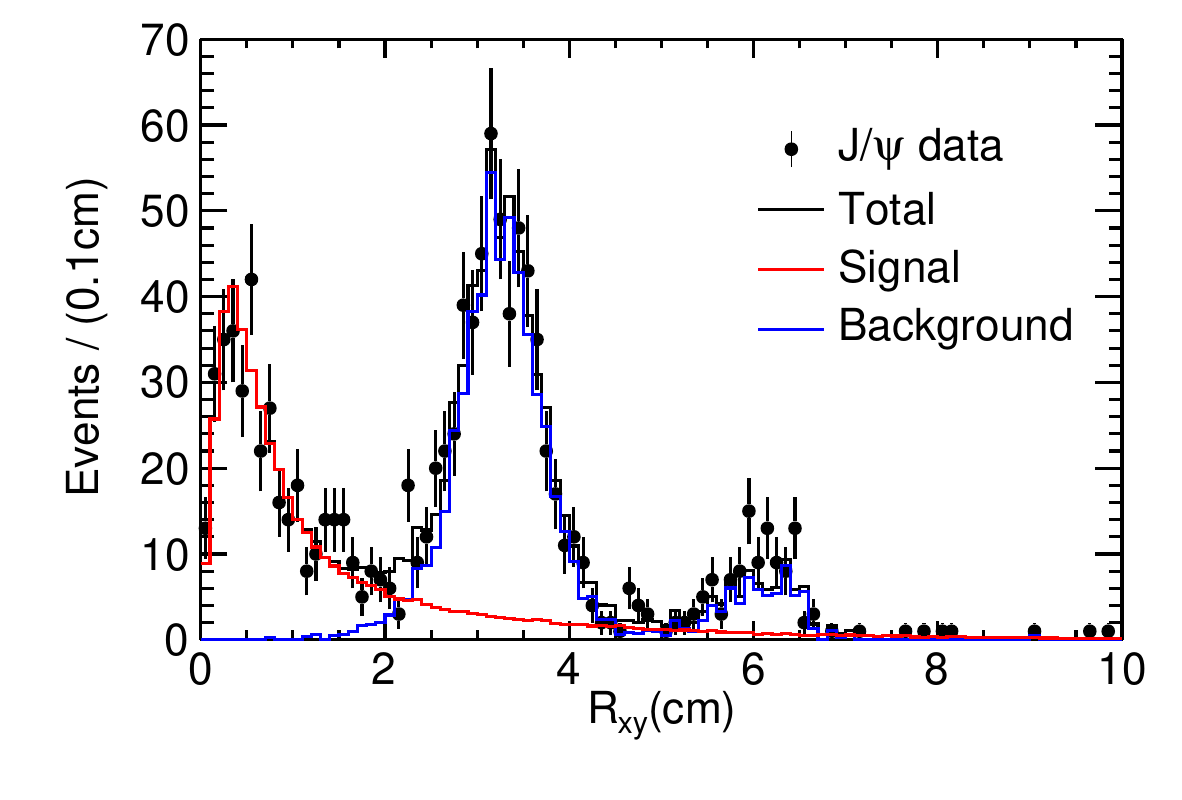}
		\end{minipage}
	}
	{
		\begin{minipage}{1.0\linewidth}
			\includegraphics[scale=0.3]{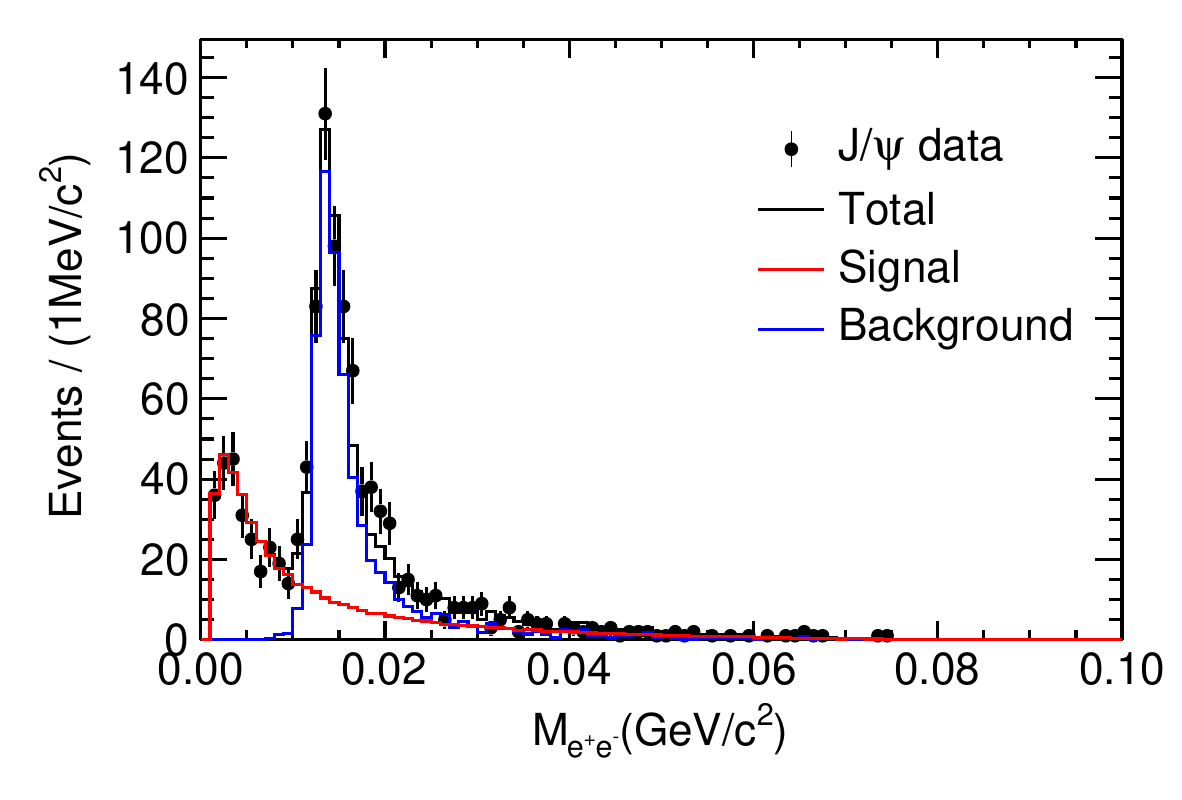}
		\end{minipage}
	}
	\caption{Distributions of (top) $R_{xy}$ and (bottom) $M_{\ee}$ from the vertex fit for $\ee$ candidates. The black dots with error bars represent data. The black, blue and red histograms are the total, gamma conversion background and signal from MC simulation, respectively.}
	\label{fig::Mee_Rxy}
    \end{center}
\end{figure}

After applying all of the above selection criteria, studies based on the inclusive MC sample indicate that the remaining background is dominated by $J/\psi \rightarrow \Sigmaa\Sigmab\rightarrow \Lambda\Lambdab\gamma\gamma$, in which a $\gamma$-conversion occurs at the beam-pipe (with an inner radius of 31.5~mm) or the MDC inner wall (with an inner radius of 54~mm).
The distribution of the radial distance between the $e^+e^-$ vertex position to the IP ($R_{xy}$) and the invariant mass of the $\ee$ ($M_{\ee}$) are shown in Fig.~\ref{fig::Mee_Rxy}, showing good separation between the signal and conversion events. To eliminate this background, a $\gamma$-conversion veto is applied based on the distributions of $M_{ee}^{BP}$~versus~$R_{xy}$ and $\Phi_{ee}$~versus~$R_{xy}$, where BP stands for beam pipe. The $\gamma$-conversion veto procedure, as well as the definitions of $M_{ee}^{BP}$ and $\Phi_{ee}$, can be found in Ref.~\cite{gammaconversionveto}.

\begin{figure}[htpb]
	\centering
    {
		\begin{minipage}{1.0\linewidth}
			\includegraphics[scale=0.4]{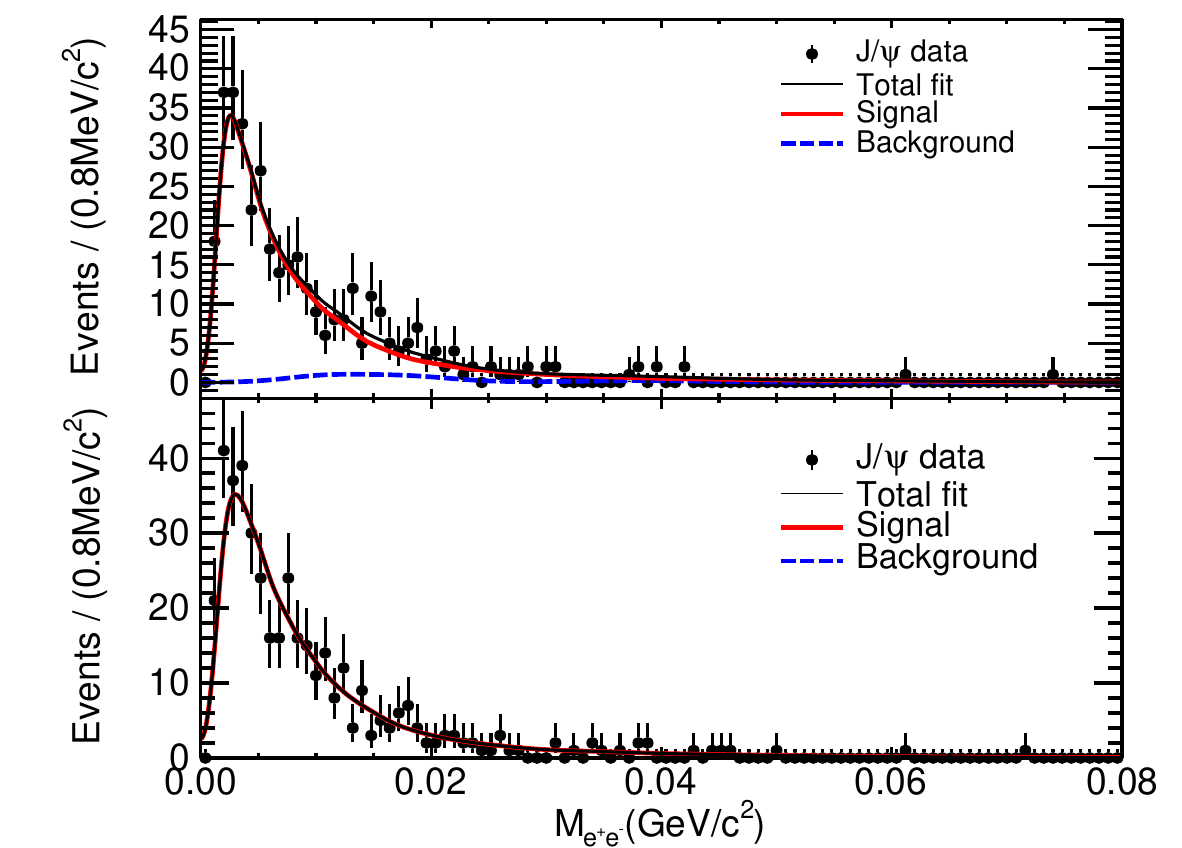}
		\end{minipage}
	}
	\caption{Fits to the $M_{\ee}$ distributions for the tagged $\Sigmab$~(top) and $\Sigmaa$~(bottom) candidates. The black dots represent data. The black, red and blue lines represent the total fits, signals and backgrounds, respectively.}
	\label{fig::Meefit}
\end{figure}

To extract DT yields, maximum likelihood fits are performed to the  $M_{\ee}$ distributions for the two charge conjugate modes separately, as shown in Fig.~\ref{fig::Meefit}.
In the fit, the components of signal and conversion background are both described by the corresponding MC simulated shapes.
The resulting DT yields and DT detection efficiencies, as well as the obtained BF of $\Siglamee$ according to Eq.~\eqref{BF-Siglamee} are summarized in Table~\ref{table::Measurement}. The fitted background yield is very close to 0, which is the lower limit when tagging the $\Sigmaa$ on the ST side. This is consistent with the expected background yield based on MC simulation within 1$\sigma$.
Studies of control samples indicate small differences in  detection efficiencies between data and MC simulation, and the corresponding correction factors
are implemented in the BF calculation.

The magnetic FF at $q^2=0$ ($G_{\textrm{M}}(0)$) is related to the measured BFs through
\begin{equation}
	\begin{aligned}
		\label{Eq::GM0withBF}
		|G_\textrm{M}(0)|^2 = \frac{\Gamma_{\textrm{total}} \mathcal{B}_{\Siglamee}}{k}
	\end{aligned}
\end{equation}
where $\Gamma_{\textrm{total}}$ is the total decay width calculated using the lifetime of the $\Sigma^0$ taken from the PDG~\cite{Particledatagroup} and $k$ is given by
\begin{equation}
	\begin{aligned}
		\label{Eq::Wq2}
		k &= \int \frac{\textrm{d}\Gamma_{\Siglamee}}{\textrm{d}q^2} \frac{1}{|G_\textrm{M}(0)|^2} dq^2 \\
		&= \int \mathrm{W}(q^2)\textrm{d}q^2
	\end{aligned}
\end{equation}
Here $\mathrm{W}(q^2)$ is independent of the form factors, therefore $k$ is fully determined with the theoretical framework~\cite{Husek2020,Batozskaya2023}.

The transition magnetic moment $\mu$ is obtained according to
\begin{equation}
	\mu =G_{\textrm{M}}(0) \frac{2 m_p}{m_{\Lambda}+m_{\Sigma}} \mu_N,
	\label{Eq::transition FFs}
\end{equation}
where $\mu_N$ represents the nucleon magnetic moment. The obtained values for $\mu$ are summarized in Table~\ref{table::Measurement}.
Comparing the results of the BF and $\mu$, good agreement is obtained between the two charge conjugate modes. Therefore, a simultaneous fit to the $M_{\ee}$ distributions between the two charge conjugate modes is performed, and the results are also summarized in  Table~\ref{table::Measurement}.

As a benefit from the DT method, all systematic uncertainties on the ST side cancel, except for those associated with the extraction of ST yields.
Assuming all sources of systematic uncertainties are uncorrelated, the total uncertainty is obtained by summing the individual contributions in quadrature, as shown in Table~\ref{table::Measurement}. Meanwhile the total correction factor is the product of all individual ones, and is implemented in the BF measurement. The systematic uncertainties are estimated as follows.

Systematic uncertainties associated with the signal $\Lambda$ reconstruction, including tracking and PID efficiencies for the proton and pion, are studied using the control sample $J/\psi~(\psi(2S)) \rightarrow pK^-\Lambdab$. Those related to the tracking of the electron and positron are evaluated using Bhabha scattering events, while uncertainties from the PID of the electron and positron  and the $\gamma$-conversion veto are studied with  $\Jpsirhopi \rightarrow \pi^+\pi^-\pi^0$, where the $\pi^0$ decays via the Dalitz channel $\pi^0\to\gamma \ee$. The $\ee$ vertex fit and the 4C kinematic fit are tested with $J/\psi \rightarrow p\bar{p}\pi^+\pi^-\pi^0$, also employing $\pi^0\to\gamma \ee$. Finally, systematic uncertainties in the extraction of ST and DT yields, due to the choice of the fitting range and the shapes of signal and background, are estimated by varying the fit range by $\pm1\sigma$ around the signal peak and by adopting alternative shape models, respectively. For the systematic uncertainty estimation of the transition magnetic moment,  the uncertainty from the $\Gamma_{\textrm{total}}$ must also be considered.
Details of the systematic uncertainties and the corresponding correction factors are summarized in the Supplemental Material~\cite{supplement}.

In summary, we present the first experimental study of the hyperon Dalitz decay $\Siglamee$, where the $\Sigma^0$ is produced in the decay $\Jpsisigsigbar$, using 10 billion $J/\psi$ events collected with the BESIII detector. The absolute BF is measured to be $ \mathcal{B}(\Siglamee) = (6.34 \pm 0.25_{\textrm{stat.}} \pm 0.23_{\textrm{syst.}}) \times 10^{-3} $, with a deviation from the PDG value ($5.5 \times 10^{-3}$) of about $2\sigma$~\cite{Particledatagroup}. The transition magnetic moment $\mu$ is measured to be $(1.74 \pm 0.03_{\textrm{stat.}} \pm 0.09_{\textrm{syst.}})\mu_N$, with a deviation within $2\sigma$ from the PDG average~\cite{Particledatagroup}. This analysis performs the most precise measurement of the  BF and transition magnetic moment, providing valuable insight into the intrinsic structure of the $\Sigma^0$ hyperon and its decay mechanism.

\input{./acknowledgement_2025-11-14.tex}

\bibliography{bibcite.bib}

\end{document}


\normalsize
\parskip=5pt plus 1pt minus 1pt

\title{
\boldmath Branching Fraction and Form Factor Measurements of the Hyperon Dalitz Decay $\Sigma^0 \rightarrow \Lambda e^+e^-$
}

\author{BESIII Collaboration}

\maketitle

\section{ST ANALYSIS}
\begin{figure}[htbp]
    \centering
	{
		\begin{minipage}{0.45\linewidth}
			\includegraphics[scale=0.35]{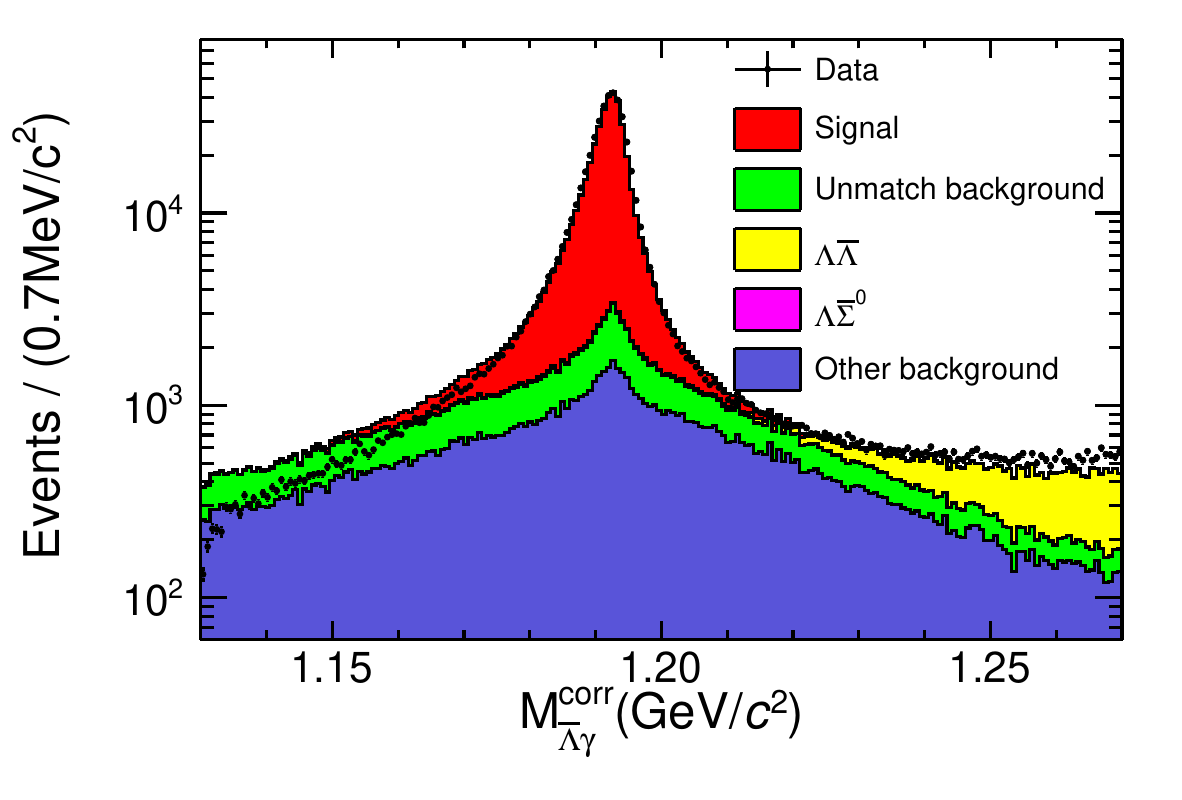}
		\end{minipage}
	}
	{
		\begin{minipage}{0.45\linewidth}
			\includegraphics[scale=0.35]{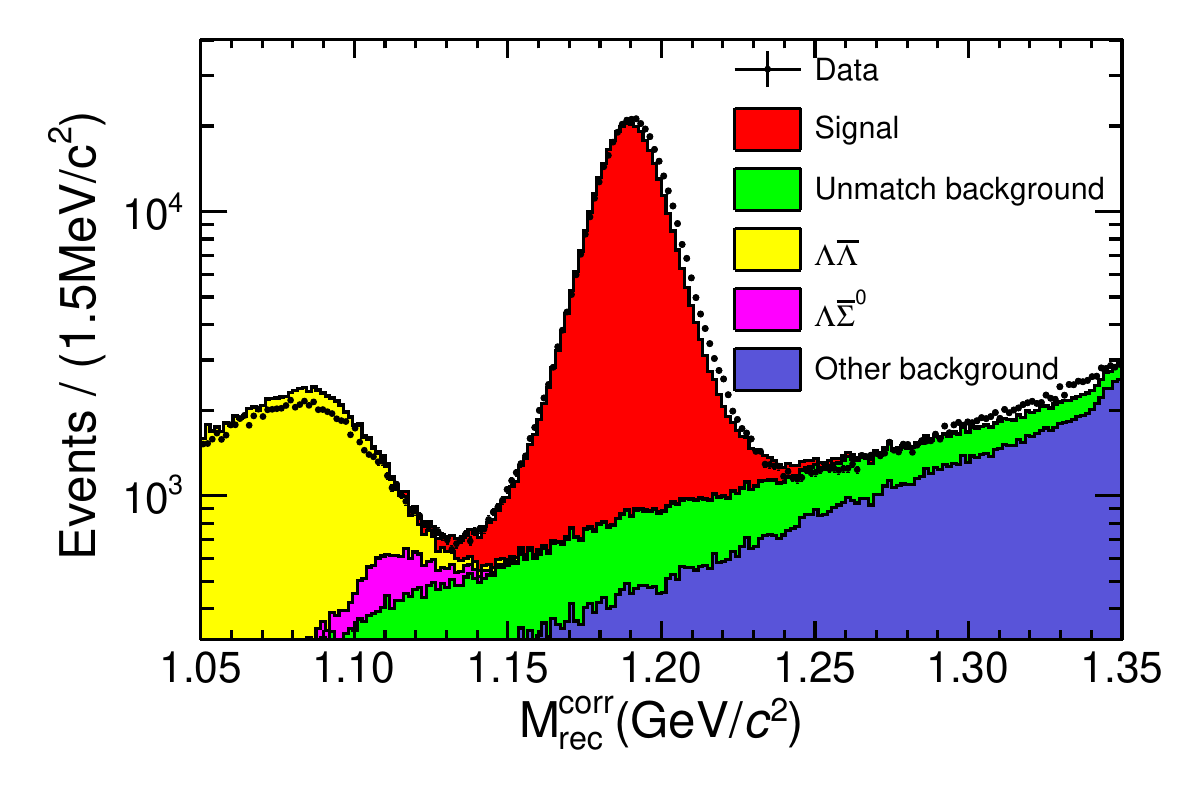}
		\end{minipage}
	}
	\caption{Distributions of $M_{\Lambdab\gamma}^{\textrm{corr}}$~(left) and $M_{\textrm{rec}}^{\textrm{corr}}$~(right) for data and the inclusive MC sample.
    The black dots represent data. The blue and green filled histograms represent the signal events with correct and fake photon selection ($\theta_{\textrm{match}} \textgreater 5^{\circ}$), the yellow and pink filled histograms represent the backgrounds with $\jpsi\to\Lambda\Lambdab$, $\jpsi\to\Lambda\Sigmab$ individually, and the red filled histogram represents the background from other processes. }
	\label{fig::1Dmasscorr}
\end{figure}

Figure~\ref{fig::1Dmasscorr} shows the distributions of $M_{\Lambdab\gamma}^{\textrm{corr}}$ and  $M_{\textrm{rec}}^{\textrm{corr}}$ for the data and inclusive MC sample, where the former one is presented without the $M_{\Lambdab\gamma}^{\textrm{corr}}$ requirement.
Good agreements between data and inclusive MC sample in the signal region of $\Jpsisigsigbar$ are observed, and the prominent peak with low background represents the $\Jpsisigsigbar$ signal in both distributions, which are separated from the truth photon with an opening angle $\theta_{\textrm{match}} \textless 5^\circ$.
The distributions of the inclusive MC sample illustrate the backgrounds come from $\jpsi\to\Lambda\Lambdab$ and $\jpsi\to\Lambda\Sigmab$ with sizable fractions, as well as a lot of other background processes with small fractions.
None of these backgrounds produce peaks in the $M_{\textrm{rec}}^{\textrm{corr}}$ distribution within the signal region of $\Sigmaa\Sigmab$.

\begin{figure}[htbp]
	\centering
    {
		\begin{minipage}{1.0\linewidth}
			\includegraphics[scale=0.4]{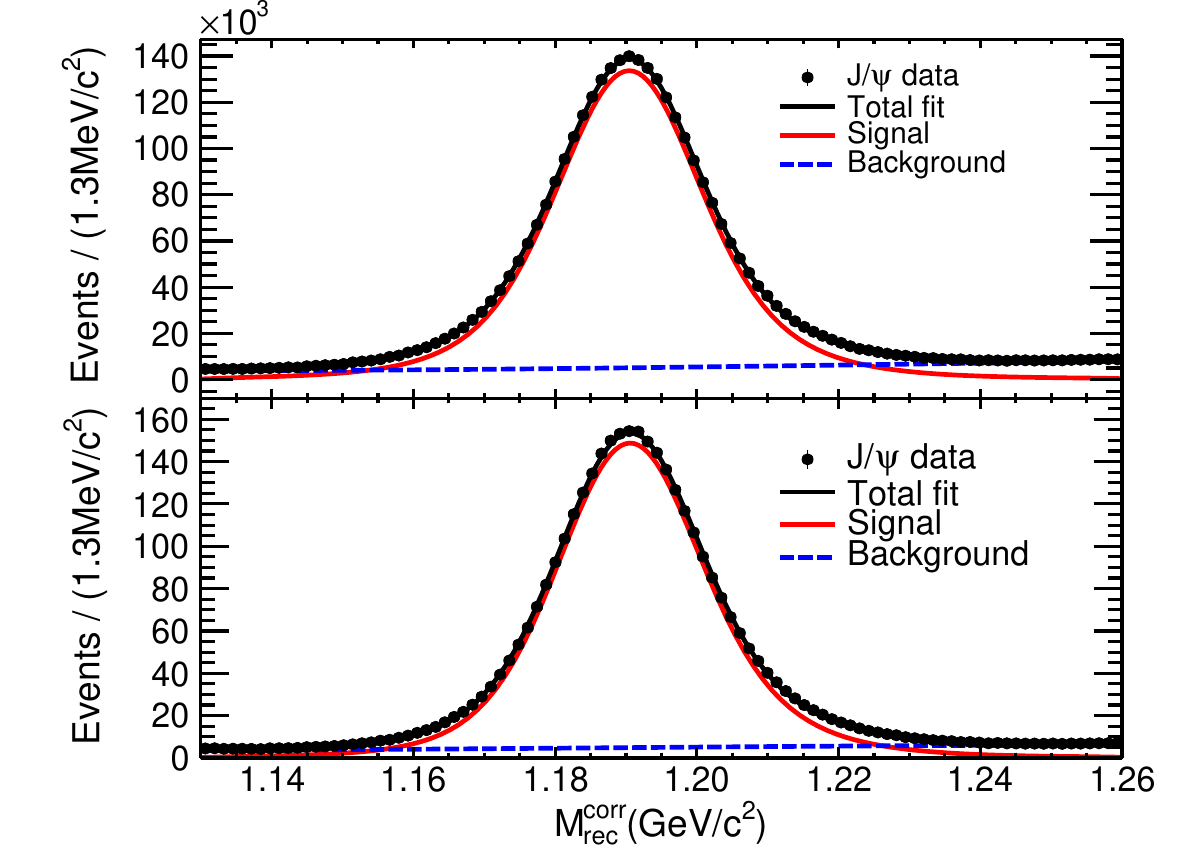}
		\end{minipage}
	}
	\caption{The fits to the $M_{\textrm{rec}}^{\textrm{corr}}$ distributions for the two conjugate channels tagged (top) $\Sigmab$ and (bottom) $\Sigmaa$. The black dots represent data. The black, red and blue curves represent the total fits, signals and backgrounds, respectively.}
	\label{fig::singletagsignal_yield}
\end{figure}

Figure~\ref{fig::singletagsignal_yield} shows the maximum likelihood fits to the $M_{\textrm{rec}}^{\textrm{corr}}$ distribution within the range between 1.13 and 1.26 GeV$/c^2$ for the two charge conjugate modes, individually.
In the fit, the signal is described with an  MC simulated shape convolved with a Gaussian function, which compensates the resolution difference between data and MC simulation. The background including unmatched $\Jpsisigsigbar$ events is described with a second order Chebyshev polynomial function.

\section{COMPARISON BETWEEN DATA AND MC SIMULATION ON DT SIDE}

\begin{figure}[htbp]
	\centering
	{
		\begin{minipage}{0.45\linewidth}
			\includegraphics[scale=0.38]{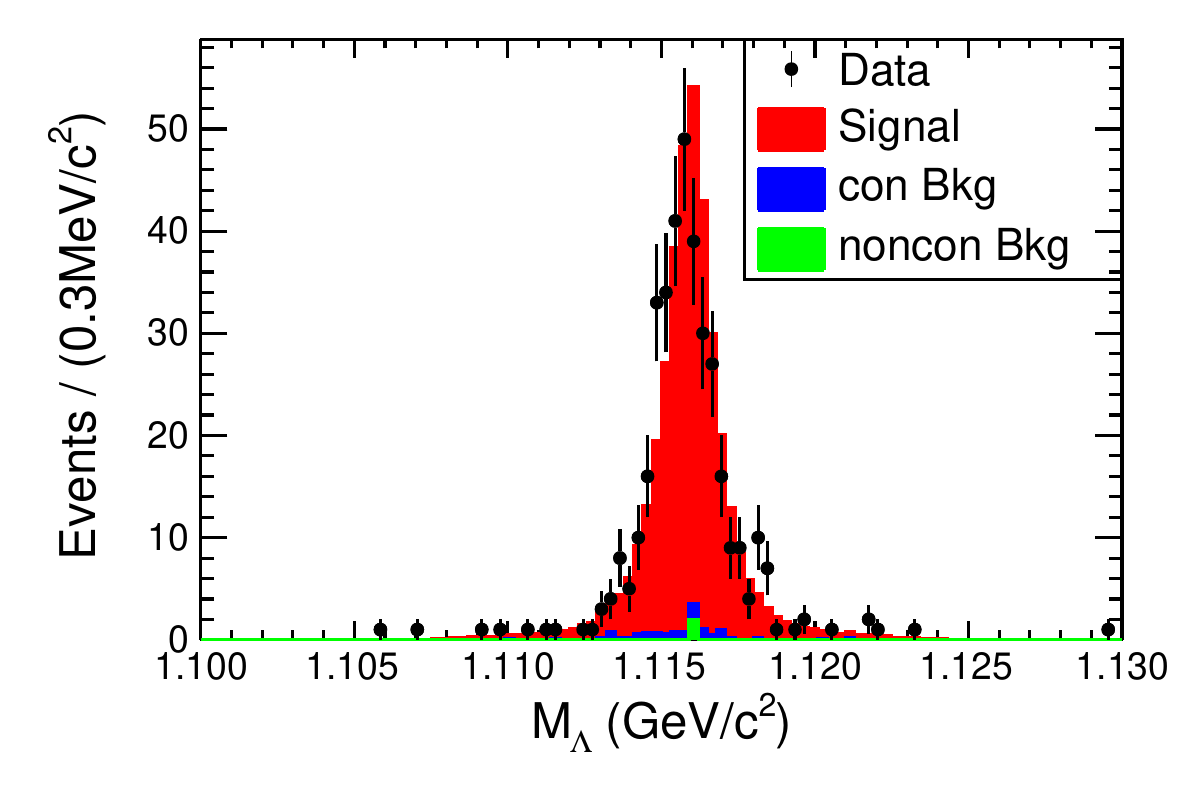}
		\end{minipage}
	}
	{
		\begin{minipage}{0.45\linewidth}
			\includegraphics[scale=0.38]{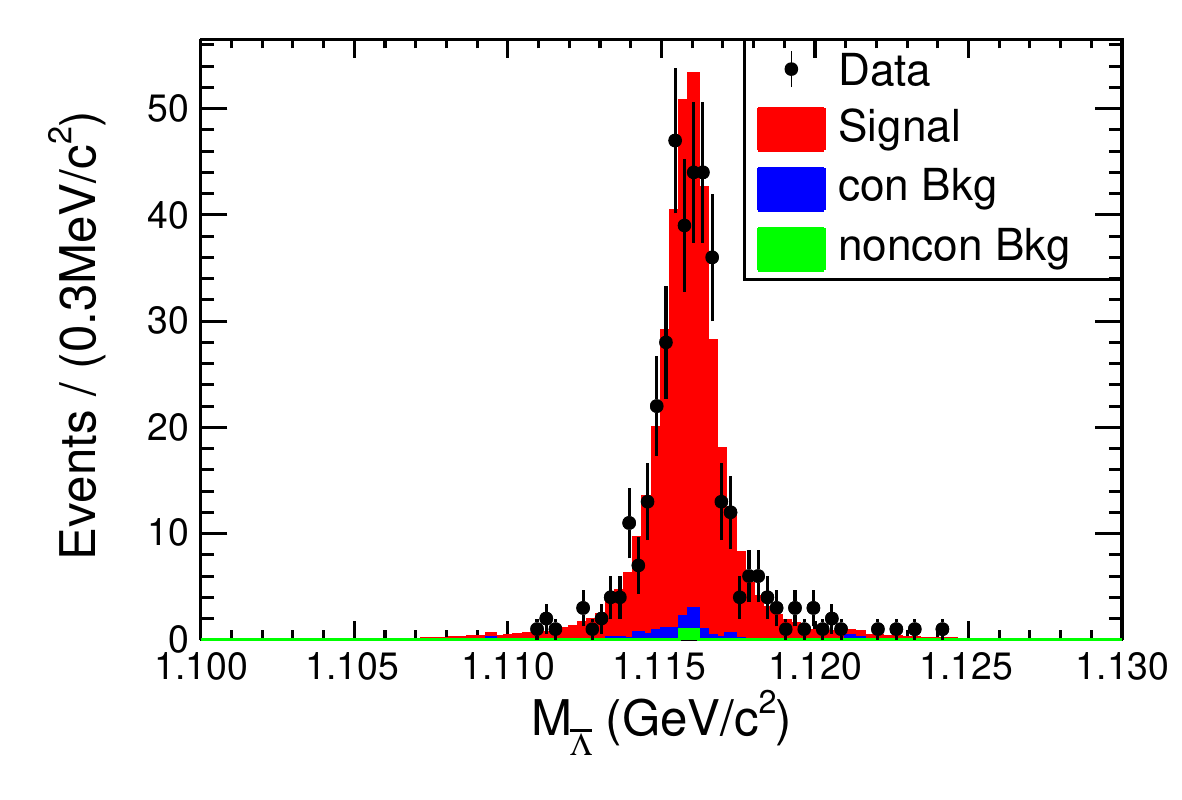}
		\end{minipage}
	}
	{
		\begin{minipage}{0.45\linewidth}
			\includegraphics[scale=0.38]{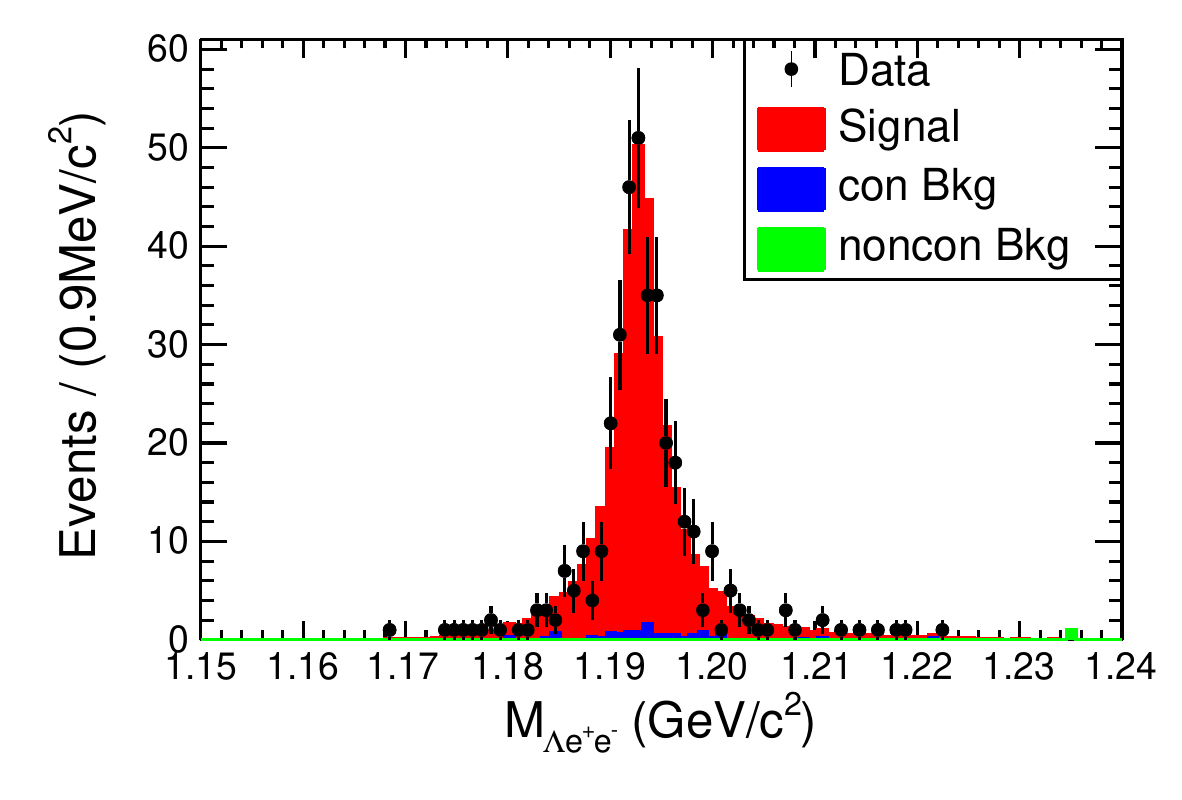}
		\end{minipage}
	}
	{
		\begin{minipage}{0.45\linewidth}
			\includegraphics[scale=0.38]{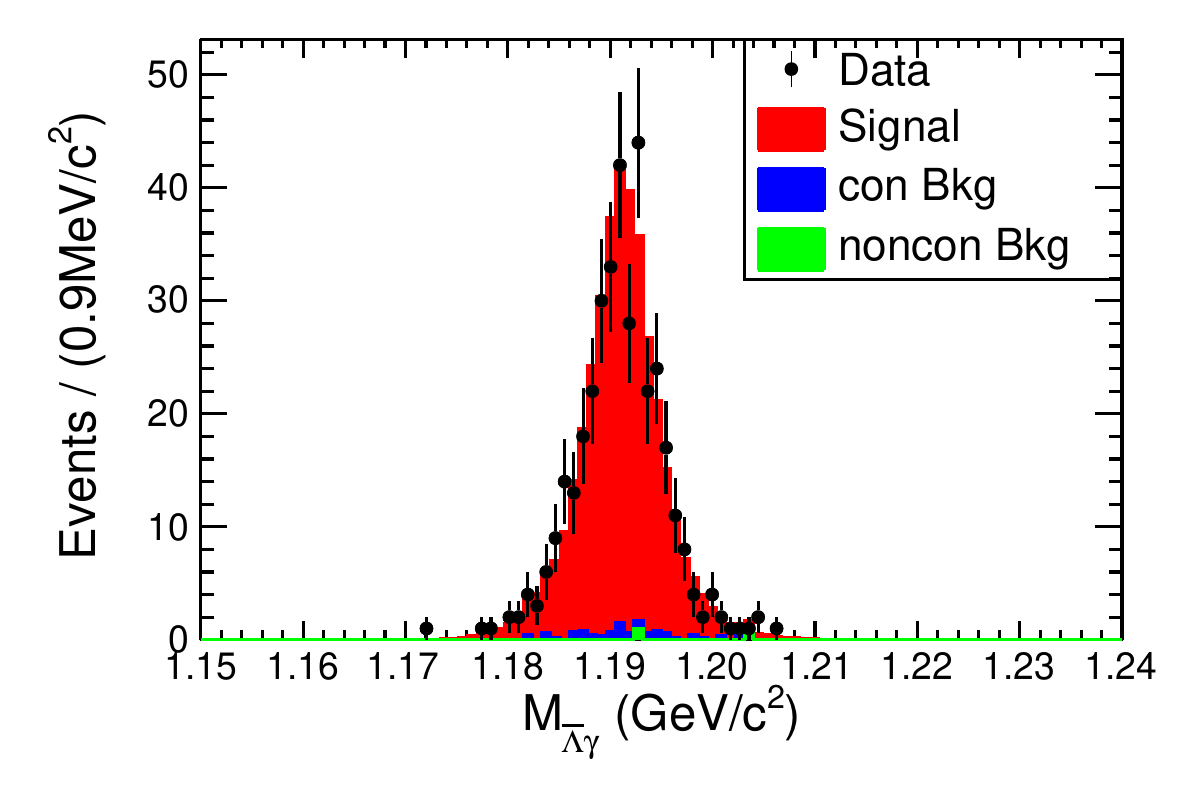}
		\end{minipage}
	}
	\caption{The invariant mass of (a) $M_{\Lambda}$, (b) $M_{\bar{\Lambda}}$, (c)  $M_{\Lambda \ee}$ and (d) $M_{\bar{\Lambda}\gamma}$ after the veto of $\gamma$-conversion. }
	\label{fig::inv mass of 4C after veto}
\end{figure}

As shown in Fig.~\ref{fig::inv mass of 4C after veto}, comparisons of various invariant mass distributions between data and simulation show good agreement. This confirms a pure sample of $\Sigmaa$ candidates on the DT side and a low remaining background level after applying the $\gamma$-conversion veto.

Due to the lack of discrimination between signal and $\gamma$-conversion background, the $M_{\Lambda \ee}$
distribution is unsuitable for yield extraction. Consequently, the more discriminating $M_{\ee}$ distribution is adopted as the spectrum for extracting the DT yields.

\section{SYSTEMATIC UNCERTAINTIES}

The systematic uncertainties of the BF measurement are from the reconstruction of $\Lambda$, tracking and PID of electron/positron, vertex fit of $\ee$, 4C kinematic fit, veto of gamma conversion as well as the extraction of ST and DT yields.
In the following studies, to account for the small difference in the selection efficiencies of charged tracks and photons between data and MC simulation, the efficiency correction factor will be applied in the BF calculation. The distributions in transverse momentum and polar angle are divided into bins. For each bin, the ratio between data and MC simulation is obtained. The average of these ratios serves as the correction factor, and its associated uncertainty is assigned as a systematic uncertainty. The combined uncertainties and correction factors are obtained by averaging those of two charge-conjugate modes.

The systematic uncertainties associated with $\Lambda$ reconstruction and electron/positron tracking and PID are studied with the control samples of $J/\psi~(\psi(2S)) \rightarrow pK^-\Lambdab$ as well as their charge-conjugate channels, where those associated with the corresponding tracking, PID as well as the vertex and secondary vertex fit are included.
The efficiencies of data and MC simulation are obtained in different $P_\textrm{T}$ versus $\cos\theta$ bins, where $P_\textrm{T}$ and $\theta$ represent the transverse momentum and the polar angle in the laboratory frame, and the ratio of weighted efficiency between data and MC simulation is calculated by incorporating the distribution of $\Lambda$ in the signal, and the corresponding mean and uncertainty are regarded as the correction factor of efficiency and systematic uncertainty, respectively.

The systematic uncertainties associated with electron/positron tracking and PID are studied with the control samples of Bhabha scattering events and $\Jpsirhopi \rightarrow \pi^+\pi^-\pi^0$ with the subsequent Dalitz decay $\pi^0\to\gamma \ee$, respectively.
The corresponding correction factors of efficiency and systematic uncertainties are obtained with the same approach as that in the study of $\Lambda$ reconstruction.

The systematic  uncertainty from the veto of gamma conversion is studied with the control sample of $\Jpsirhopi \rightarrow \pi^+\pi^-\pi^0$ with the  subsequent Dalitz decay $\pi^0\to\gamma \ee$ and with the additional requirements of the invariant mass $|M_{\gamma\ee} - M_{\pi^0}| \textless 20$~MeV$/c^2$ and the $\pi^+\pi^-$ recoiling mass $M_{\pi^+\pi^-}^{\textrm{rec}} \textless 300$~MeV$/c^2$. The gamma conversion algorithm and the same veto requirement are implemented on the control sample. The resultant difference on the efficiencies between data and MC simulation is  taken as the systematic uncertainty.

The systematic uncertainty associated with the vertex fit of $\ee$ and 4C kinematic fit is studied with the control sample of $J/\psi \rightarrow p\bar{p}\pi^+\pi^-\pi^0$ with the subsequent Dalitz decay $\pi^0 \rightarrow \gamma\ee$. The same requirements on $M_{\gamma\ee}$ and $M_{\pi^+\pi^-}^{\textrm{rec}}$ as those in the study of gamma conversion veto are implemented and the resultant difference of the efficiencies between data and MC simulation is taken as the systematic uncertainty.

The systematic uncertainty from ST yield extraction comes from the fitting range and signal shape. The systematic uncertainty related to the fitting range is estimated by the alternative fits via varying the fitting range by $\pm$0.01~GeV$/c^2$, the larger change of ST yield is taken as the systematic uncertainty. The systematic uncertainty related to the signal shape is estimated with the alternative fit by varying the width of convoluted Gaussian function within its uncertainty, and no obvious difference in the signal yield is observed. The systematic uncertainty related to the background shape is estimated by changing the polynomial function to the simulated background shape, which includes the inclusive MC processes without $\Jpsisigsigbar$ and unmatched MC scaled to data. Finally no obvious difference is observed in both cases.

The systematic uncertainty from DT yield is estimated by comparing alternative fit strategies. For signal MC, we convolve the signal MC shape with a Gaussian resolution function, and finally no obvious difference is observed. For background MC, there is no better way to accurately describe background process, so we estimate it with the difference between the nominal fitting result and the result when we fix the background level to expectation from MC simulation.

For the systematic uncertainty of the transition magnetic moment, the contribution arising from the uncertainty in the $\Sigmaa$ total decay width is estimated to be 4.7\%. The width is calculated from the $\Sigma^0$ mean lifetime, $(7.4 \pm 0.7) \times 10^{-20}$~sec, as quoted from the PDG.

Table~\ref{table::Sys} summarizes all the systematic uncertainties.

Table~\ref{table::Corr} summarizes all the correction factors.

\begin{table*}[htpb]
	\centering
	\renewcommand\arraystretch{1.2}
	\caption{Summary of the systematic uncertainties in the BF measurement.}
	\label{table::Sys}
	\begin{tabular}{l c c c}
		\hline
		\hline
		Source    & ~~$\Siglamee$~(\%)~~ & ~~$\Sigblamee$~(\%)~~ & ~~Combined~(\%)~~ \\
		\hline
		$\Lambda(\Lambdab)$ reconstruction       &  1.0  & 1.0 & 1.0  \\
		$e^{+}$ tracking       &   0.3  &  0.3  &  0.3  \\
		$e^{-}$ tracking       &   0.4  &  0.4  &  0.4  \\
		$e^{+}$ PID      &   0.3  &  0.3  &  0.3  \\
		$e^{-}$ PID      &   0.3  &  0.3  &  0.3  \\
		Gamma conversion veto      &   0.2  &  0.2  &  0.2  \\
		$e^{+}e^{-}$ vertex fit and 4C kinematic fit    &   1.3  &  1.3  &  1.3  \\
		ST fit      &   1.5  &  0.6  &  1.1    \\
		DT fit      &   2.8  &  2.4  &  2.6    \\
		\hline
		Total   &  3.8  &  3.3  &  3.5  \\
		\hline
		\hline
	\end{tabular}
\end{table*}

\begin{table*}[htpb]
	\centering
	\renewcommand\arraystretch{1.2}
	\caption{Summary of the correction factors in the BF measurement.}
	\label{table::Corr}
	\begin{tabular}{l c c}
		\hline
		\hline
		Source    &  ~~~$\Siglamee$~(\%)~~~ & ~~~$\Sigblamee$~(\%)~~~  \\
		\hline
		$\Lambda(\Lambdab)$ reconstruction       &  1.04  &  1.01   \\
		$e^{+}$ tracking       &    1.01  &  1.01   \\
		$e^{-}$ tracking       &    1.04  &  1.04   \\
		$e^{+}$ PID       &    0.95  &  0.95   \\
		$e^{+}$ PID       &    0.93  &  0.93   \\
		\hline
		Total             &   0.97  &  0.94   \\
		\hline
	\end{tabular}
\end{table*}

%% file: authorlist_2025-11-14.tex
M.~Ablikim$^{1}$\BESIIIorcid{0000-0002-3935-619X},
M.~N.~Achasov$^{4,c}$\BESIIIorcid{0000-0002-9400-8622},
P.~Adlarson$^{82}$\BESIIIorcid{0000-0001-6280-3851},
X.~C.~Ai$^{88}$\BESIIIorcid{0000-0003-3856-2415},
C.~S.~Akondi$^{31A,31B}$\BESIIIorcid{0000-0001-6303-5217},
R.~Aliberti$^{39}$\BESIIIorcid{0000-0003-3500-4012},
A.~Amoroso$^{81A,81C}$\BESIIIorcid{0000-0002-3095-8610},
Q.~An$^{78,64,\dagger}$,
Y.~H.~An$^{88}$\BESIIIorcid{0009-0008-3419-0849},
Y.~Bai$^{62}$\BESIIIorcid{0000-0001-6593-5665}
% ...
\\
\vspace{0.2cm}
(BESIII Collaboration)\\
\vspace{0.2cm} {\it
$^{1}$ Institute of High Energy Physics, Beijing 100049, People's Republic of China\\
% ...
$^{4}$ Budker Institute of Nuclear Physics SB RAS (BINP), Novosibirsk 630090, Russia\\
% ...
$^{31}$ INFN Sezione di Ferrara, (A)INFN Sezione di Ferrara, I-44122, Ferrara, Italy; (B)University of Ferrara, I-44122, Ferrara, Italy\\
% ...
$^{39}$ Johannes Gutenberg University of Mainz, Johann-Joachim-Becher-Weg 45, D-55099 Mainz, Germany\\
% ...
$^{62}$ Southeast University, Nanjing 211100, People's Republic of China\\
% ...
$^{64}$ State Key Laboratory of Particle Detection and Electronics, Beijing 100049, Hefei 230026, People's Republic of China\\
% ...
$^{78}$ University of Science and Technology of China, Hefei 230026, People's Republic of China\\
% ...
$^{81}$ University of Turin and INFN, (A)University of Turin, I-10125, Turin, Italy; (B)University of Eastern Piedmont, I-15121, Alessandria, Italy; (C)INFN, I-10125, Turin, Italy\\
$^{82}$ Uppsala University, Box 516, SE-75120 Uppsala, Sweden\\
% ...
$^{88}$ Zhengzhou University, Zhengzhou 450001, People's Republic of China\\

\vspace{0.2cm}
$^{\dagger}$ Deceased\\
% ...
$^{c}$ Also at the Novosibirsk State University, Novosibirsk, 630090, Russia\\
% ...

}

%% file: acknowledgement_2025-11-14.tex
\textbf{Acknowledgement}

The BESIII Collaboration thanks the staff of BEPCII (https://cstr.cn/31109.02.BEPC) and the IHEP computing center for their strong support. This work is supported in part by Beijing Natural Science Foundation of China (BNSF) under Contract No. IS23014; National Key R\&D Program of China under Contracts Nos. 2025YFA1613900, 2023YFA1606000, 2023YFA1606704; National Natural Science Foundation of China (NSFC) under Contracts Nos. 11635010, 11935015, 11935016, 11935018, 12025502, 12035009, 12035013, 12061131003, 12165022, 12192260, 12192261, 12192262, 12192263, 12192264, 12192265, 12221005, 12225509, 12235017, 12342502, 12361141819; the China Postdoctoral Science Foundation under Grant No. 2024M753040; the Postdoctoral Fellowship Program of China Postdoctoral Science Foundation under Grant No. GZC20241608; the Chinese Academy of Sciences (CAS) Large-Scale Scientific Facility Program; the Strategic Priority Research Program of Chinese Academy of Sciences under Contract No. XDA0480600; CAS under Contract No. YSBR-101; 100 Talents Program of CAS; The Institute of Nuclear and Particle Physics (INPAC) and Shanghai Key Laboratory for Particle Physics and Cosmology; Yunnan Fundamental Research Project under Contract No. 202301AT070162; ERC under Contract No. 758462; German Research Foundation DFG under Contract No. FOR5327; Istituto Nazionale di Fisica Nucleare, Italy; Knut and Alice Wallenberg Foundation under Contracts Nos. 2021.0174, 2021.0299, 2023.0315; Ministry of Development of Turkey under Contract No. DPT2006K-120470; National Research Foundation of Korea under Contract No. NRF-2022R1A2C1092335; National Science and Technology fund of Mongolia; Polish National Science Centre under Contract No. 2024/53/B/ST2/00975; STFC (United Kingdom); Swedish Research Council under Contract No. 2019.04595; U. S. Department of Energy under Contract No. DE-FG02-05ER41374